\documentclass[letterpaper, 10 pt, conference]{ieeeconf}  

\IEEEoverridecommandlockouts                              

\usepackage{graphicx} 
\usepackage{epsfig} 
\usepackage{amsmath} 

\title{\LARGE \bf
Cluster-Span Threshold: An Unbiased Threshold for Binarising Weighted Complete Networks in Functional Connectivity Analysis
}

\author{Keith Smith$^{1,2}$, \textit{student member, IEEE}, Hamed Azami$^{1}$, \textit{student member, IEEE},\\ Mario A. Parra$^{2}$, John M. Starr$^{2}$, Javier Escudero$^{1}$, \textit{member, IEEE}
\thanks{*This work was partially supported by the Engineering and Physical Sciences Research Council. MAP was awarded an MRC Centenary Early Career Award \#MRC-R42552. This work was conducted within the context of The University of Edinburgh Centre for Cognitive Ageing and Cognitive Epidemiology, part of the cross council Lifelong Health and Wellbeing Initiative (MR/K026992/1). MAP work is currently supported by Alzheimer's Society, Grant \#AS-R42303.}
\thanks{$^{1}$Keith Smith, Hamed Azami and Javier Escudero are with the Institute for Digital Communications, School of Engineering,
        University of Edinburgh,  King's Buildings, West Mains Road, Edinburgh, UK, EH9 3JL.
        {\tt\small k.smith@ed.ac.uk, hamed.azami@ed.ac.uk, javier.escudero@ed.ac.uk }}%
\thanks{$^{2}$Keith Smith, John M. Starr and Mario A. Parra are with the Alzheimer Scotland Dementia Research Centre, School of Philosophy, Psychology and Language Sciences, University of Edinburgh
		7 George Square, Edinburgh, EH8 9JZ.
        {\tt\small jstarr@staffmail.ed.ac.uk, mprodri1@staffmail.ed.ac.uk}}%
}

\begin{document}

\maketitle
\thispagestyle{empty}
\pagestyle{empty}

\begin{abstract}

We propose a new unbiased threshold for network analysis named the Cluster-Span Threshold (CST). This is based on the clustering coefficient, $C$, following logic that a balance of `clustering' to `spanning' triples results in a useful topology for network analysis and that the product of complementing properties has a unique value only when perfectly balanced. We threshold networks by fixing $C$ at this balanced value, rather than fixing connection density at an arbitrary value, as has been the trend. We compare results from an electroencephalogram data set of volunteers performing visual short term memory tasks of the CST alongside other thresholds, including maximum spanning trees. We find that the CST holds as a sensitive threshold for distinguishing differences in the functional connectivity between tasks. This provides a sensitive and objective method for setting a threshold on weighted complete networks which may prove influential on the future of functional connectivity research.

\end{abstract}

\section{Introduction}
Network theory is an applied form of graph theory which is an abstract framework for studying topologies consisting of a set of nodes with connections formed between them. Network theory is widely applied to study the complex, interdependent nature of real world phenomena \cite{c11} and is increasingly used for the analysis of functional brain recordings \cite{c14}\cite{c15}. 

Indeed, it seems naturally suited to applications in brain connectivity analysis. Recordings obtained from the brain, using functional magnetic resonance imaging (fMRI), the electroencephalogram (EEG) and the magnetoencephalogram (MEG) for instance, can be processed into networks via some similarity or dependency measure applied pair-wise in either the frequency or time domain. The nodes of the network can be easily defined. In the EEG and MEG we can take a 1-1 mapping between nodes and sensors whereas slightly more involved techniques can be used to define nodes from  fMRI recordings \cite{c21}. For a network with $n$ nodes, we obtain an $n$x$n$ weighted adjacency matrix. The $i$th row and $j$th column element of this matrix corresponds to the value of the similarity or dependency measure applied between nodes $i$ and $j$ \cite{c11}. Weighted connections exist between each possible pair of nodes with magnitudes between $0$ and $1$, thus the related networks cannot be distinguished by topology of existing connections, as is easily evaluated in networks \cite{c11}, but instead by the relative strengths of connections. Subjecting the matrix values to a threshold is then desirable as there are many spurious low weights and it conforms the networks to a simplified and well understood form without too much loss of information. However, until now, an objective threshold has yet to be proposed. This results in arbitrary study-to-study choices leading to different and sometimes conflicting results \cite{c3}. 

Illustrating this issue, measurements of network topologies are generally strongly dependent on the connection density of the network. Two examples of this are the most widely used measures, namely, the clustering co-efficient, $C$, and the characteristic path length, $L$. For instance, given a connected simple graph, it is straightforward to prove that the following equivalence holds:

\begin{equation*}\label{eq:smallworld}
C = 1 \iff \text{complete graph} \iff L = 1,
\end{equation*}
where $1$ is the maximum value of $C$ and the minimum value of $L$. Thus, it  can be expected that different thresholds result in different possible ranges of measurement values, confounding comparisons between studies. 

The dualism of integration-segregation plays a major role in the discussion of network topology. It is seen that brain networks find a trade-off between integrational and segregational activity \cite{c27}, which we refer to as small-world networks \cite{c26}. Segregation is seen to be represented by $C$ and integration by $L$ \cite{c23}\cite{c24}. The small-world property is then $\sigma = (C/C_{ran})/(L/L_{ran})$ where $C_{ran}$ and $L_{ran}$ are the values for the random graph ensemble with the same number of nodes and edges, normalizing $\sigma$. 

We note here that the framework of networks itself is a generic mathematical tool free from concepts of the field in which it is applied. We aim to look at network topology from this generic perspective. Although $\sigma$ is used to find where there is a balance between integration and segregation, it is not mathematically clear what this property evaluates. Further, integration and segregation are two sides of the same coin so using two heretofore unconsolidated measures to explain them seems unnecessary. Given this, we focus on clustering in graphs to come to a concise metric in order to understand network topology of weighted complete graphs more plainly. In doing so we propose a suitable non-arbitrary threshold for weighted complete networks.


Triples play a key role in understanding topology. They are formed by two 'neighbouring' nodes sharing connections with one common node. A triple forms a triangle when the neighbouring nodes are also connected. The clustering co-efficient, then, is defined as the probability that a triple is part of a triangle. It is these triangles which are seen particularly representative of segregational activity. However, the triples which do not form part of a triangle, `spanning' triples, should be deemed equally important to clustering triples in a given network since it is the relationship between clustering and spanning triples which determines in a large part the topology of the graph. Therefore, accepting $C$ as a useful measure of segregation, we propose that integration can be represented by `spanning' triples. Thus, a scale of integration-segregation may be more simply understood using the single property of the ratio of `clustering' to `spanning' triples, rather than the more obscure relationship between $C$ and $L$. 

Following from this, we hypothesise that a balance of clustering to spanning triples is a desirable topology to study. Thus, instead of looking to threshold networks based on how complete the graph is, it may instead be useful to threshold networks when this balance is reached.

The cluster-span threshold (CST), introduced here, is defined by the balance of triples which `cluster' to the triples which `span'. Here we show our first results using the CST. We compare these results with maximum spanning trees (MSTs)- a promising method which negates thresholds \cite{c5}\cite{c4}, as well as standard connection density thresholds. Particularly, we are interested in whether or not the CST is a sensitive tool for detecting differences in EEG activity of young, healthy adults during a cognitive task related to the specific detection of Alzheimer's disease \cite{c1}\cite{c2}.

\section{Materials}

\subsection{Subjects}

EEG signals were recorded for 23 healthy young volunteers participating in different Visual Short-Term Memory (VSTM) tasks. Of the volunteers, five were left-handed and eight were women. Written consent was given by all subjects and the study was approved by the Psychology Research Ethics Committee, University of Edinburgh.

\subsection{Tasks} 

\begin{figure}[thpb]
	\centering
	\includegraphics[scale=0.19]{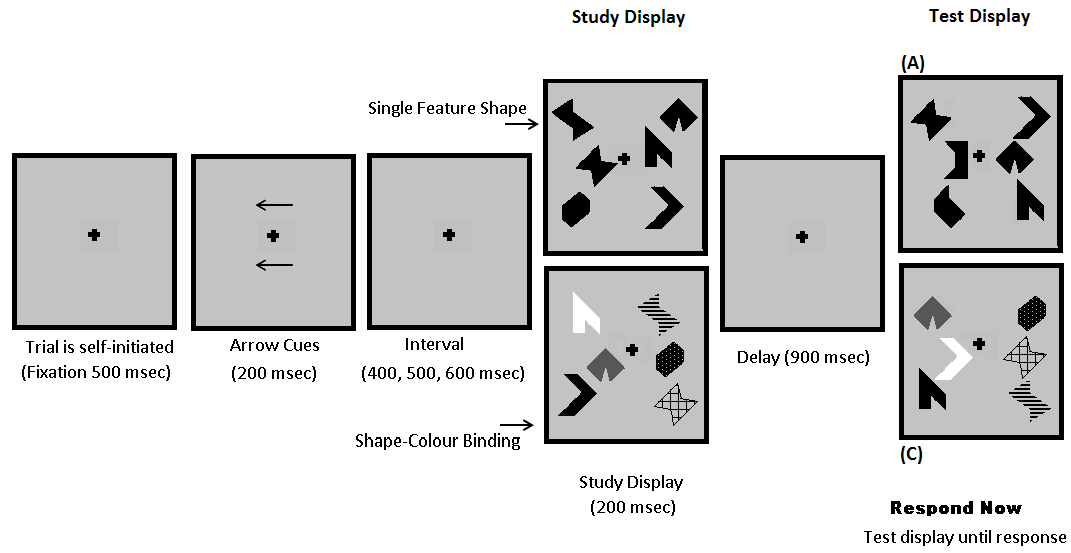}
	\caption{Structure of the VSTMB tests}
	\label{Taskfig}
\end{figure}

A schematic diagram of the test is shown in Fig. \ref{Taskfig} which also gives examples of the uncommon shapes. Three objects were presented in both sides of the screen. The patients had been signalled that the test was for the objects in the left side of the screen. Due to contralateral behaviour, this refers to right hemisphere response (RHPR). The two
tasks were distinguished by number of features of the objects. One of the tasks involved memorising single feature objects consisting of different shapes, while the other task involved memorising objects with both different shapes and different colours.

The positions of the objects were randomised separately for study and test displays to ensure that position was not a factor in memorisation. Participants completed 8 practice trials followed by 170 test trials for each task. Participants were tested on whether or not the objects in the study display were the same as the test display, which the volunteers indicated by pressing buttons with both hands. In half of the trials the objects were the same and in the other half they were different. 

\subsection{Recordings}

Trials with serious artefacts or incorrect responses were discarded since these would not inform meaningfully on memory binding activity. Problems in a few cases meant only data for 20 volunteers in the shape only task and 21 in the shape-colour binding task was available for comparisons. 

We analysed the working memory representation, consisting of study display and maintenance periods, since this is found to be where differences in brain activity are most prevalent \cite{c19}. The EEG data consisted of epochs with length of 1 second with -0.2 seconds of pre-stimulus recordings recorded in NeuroScan version 4.3 at 250 samples per second. Forty EEG channels were recorded from common EEG sites, the majority of which were international 10/20 sites. Only thirty channels were kept for our purposes. The ten discarded channels consisted of four ocular channels, two linked mastoid reference channels and four which were discarded due to systematic noise (T5, T6, FT9 \& FT10).

\section{Methods}

\subsection{Signal Processing}

Pre-processing, frequency analysis and connectivity analysis were performed using FieldTrip \cite{c6}. Channels were re-referenced using an average reference. The muli-taper method was applied from 0 seconds onwards using Slepian sequences with 2 Hz spectral smoothing. A 0.5 Hz resolution was acquired with one second of zero-padding. The data was partitioned into five frequency bands of which we focus here on $\beta$ (12.5-32 Hz) which is important in integration of senses and memory \cite{c22}. The debiased weighted phase-lag index \cite{c7} was used to obtain connectivity matrices which were average over trials. This similarity measure was chosen because it is robust to volume conduction effects and measures time-lagged dependencies between signals \cite{c16}. 

\subsection{Proposal of New Cluster-Span Threshold}

Taking 'triangles' to mean the number of clustering triples in a graph, the global clustering co-efficient is defined as follows:
\[
C = \frac{\text{triangles}}{\text{triples}}.
\]

It is known that $C$ depends strongly on the connection density, $P$, of the graph. This can be shown deductively in the case of the ensemble of Erd\"os-R\'enyi Random graphs \cite{c20}. We define $n$ as the number of nodes in a graph and $m$ as the number of connections. Then we have

\[
P = \frac{2m}{n(n-1)}. 
\]

This is also the probability that any connection exists. In this graph ensemble, $C$ forms a strictly linear relationship with the number of connections in the graph, $E[C]=P$ \cite{c11}. Furthermore, a decreasing relationship is seen with increasing threshold in functional brain networks \cite{c17}. 

However, we note that $C$ acts well as a topological measure of networks under a fixed threshold, where values are found to be different for diseased and healthy brains for instance \cite{c3}\cite{c25}. Equally, it must be that if we fix the level of clustering in a network, differences will be shown in the connection density of these networks. Therefore, an unbiased method of choosing the level of clustering can work as an unbiased way of applying a threshold to networks.

We hypothesise that a more balanced level of clustering triples to spanning triples may be desirable for network analysis. Thus, we look to the balance between the number of triangles in the graph to the number of non-triangle triples. That is, the property

\[
\text{triangles} \times \text{non-triangle triples}.
\]
Realising that the number of non-triangle triples is just the number of triples minus the number of triangles, we get

\[
\text{triangles} \times (\text{triples} - \text{triangles}).
\]
To normalise this value we use the simple mathematical property that for $a>b$, $b(a-b)$ attains its maximum value at $b = a/2$. This gives 
\[
\begin{array}{ccl}
	
	C_{B} & =  & \frac{\text{triangles} \times (\text{triples} - \text{triangles})}{\left(\frac{\text{triples}}{2}\right)^{2}}\\
	 & =  & 4\times\left(\left( \frac{\text{triangles}}{\text{triples}}\right)
	- \left(\frac{\text{triangles}}{\text{triples}}\right)^{2} \right)\\
	 & = & 4C(1-C).\\
\end{array}
\]
We call $C_{B}$ the balanced clustering co-efficient. 

\begin{figure}[thpb]
	\centering
	\includegraphics[scale=0.5]{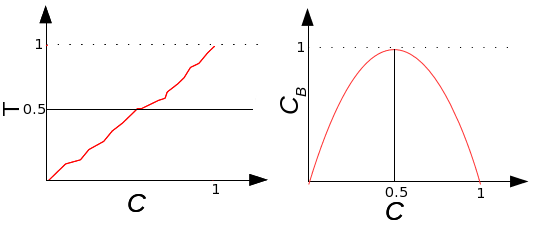}
	\caption{Simplified diagram of behaviour of $C$ against proportion of strongest connections kept in functional brain networks, $T$, and against $C_{B}$. }
	\label{Simplefig}
\end{figure}

The utility of this property can be seen in that it attains only one unique value at exactly $C=0.5$ (see Fig. \ref{Simplefig}). From this value we obtain the corresponding threshold value which we call the cluster-span threshold (CST).

In application, although generally $C$ decreases as the connection density decreases, it can still be the case that at a local level, the graph of $C$ against threshold crosses $0.5$ more than once. Also, It is rarely exactly $0.5$ because of the discrete nature of connection density. In order to take account of these details in our study, we computed the binarised networks for a suitably wide range of connection densities and chose the networks for which the number of strongest connections present obtains the value of $C$ nearest to $0.5$.   

\subsection{Comparison Networks}

Binarised networks computed from thresholds keeping the 20\%, 30\% and 40\% of strongest connections, rounded to the nearest integer, were obtained for comparisons. We also computed MSTs, processed in an identical fashion, for comparison.

\subsection{Network Measures}

We computed  $L$ and $P$ for the CST, and $L$ and $C$ for 20\%, 30\% and 40\% threshold binarised networks. As well as this we use findings of the leaf fraction and diameter of MSTs. 

The diameter (\textit{Di}) of an MST is the longest shortest path between pairs of nodes, which is linked to $L$. The leaf fraction, (\textit{LF}) is the fraction of 1-degree nodes in the MST, which is linked to $C$ \cite{c5}.  Because the CST fixes the value of clustering in the network rather than connections, this enables the use of connection density, \textit{P}, for analysis. This is simply the number of connections present in the network divided by the number of connections in the complete binary network with the same number of nodes.

All measures were computed either using the Brain Connectivity Toolbox \cite{c13}, or using simple computations in MATLAB. 

\section{Results and Discussion}
The average connection density for the CST networks in this analysis was $37.67\pm 4.38 \%$. As an example of suitability, this lies within a desirable range for brain pathology \cite{c25}. This mean breaks down into $39.06\pm 2.78\%$ for shape only and $36.20\pm 5.27\%$ for shape-colour binding tasks. In comparison, the ensemble of Erd\"os-R\'enyi random graphs has $P=C= 50\%$, which is far larger. 

\begin{table}[h]
	\caption{\textit{p}-values for the paired \textit{t}-test of Shape only vs. Shape-Colour Binding conditions in right hemisphere response.}
	\label{PTTTable}
	\begin{center}
		\begin{tabular}{|c||c||c||c||c||c|}
			\hline
			Metric	 	& CST 		& 40\% 		& 30\% 		& 20\% 		& MST\\
			\hline
			\textit{P}	 		& 0.0148	& - 		& - 		& - 		& - \\
			$C$			& -			& 0.0330	& 0.0236 	& 0.3174 	& - \\
			\textit{LF}			& - 		& - 		& - 		& -			& 0.7748\\
			$L$			& 0.0267	& 0.7233 	& 0.7864 	& 0.0243	& - \\
			\textit{Di}			& - 		& - 		& - 		& -			& 0.1036\\
			\hline
		\end{tabular}
	\end{center}
\end{table}

\begin{figure}[thpb]
	\centering
	\includegraphics[scale=0.29]{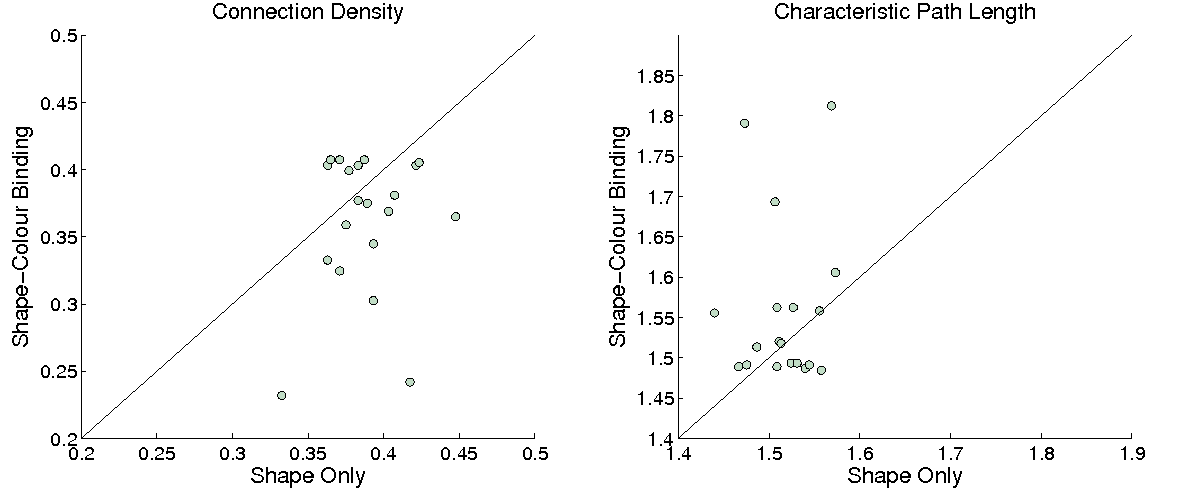}
	\caption{Scatter plots of CST networks: shape only values vs. shape-colour binding values for connection density, left, and characteristic path length, right.}
	\label{Scatterfig}
\end{figure}

Paired \textit{t}-test values were computed for the metrics for shape only vs. shape-colour binding tasks. Table \ref{PTTTable} reports the \textit{p}-values for each metric applied to each type of network.

The \textit{p}-values show that the CST works as a sensitive tool for the detection of shape only vs. shape-colour binding task differences in right hemisphere response. \textit{P} gives significant difference in the CST networks, where $C$ gives significant difference in the 40\% and 30\% thresholds. Yet, \textit{LF} fails to find this difference in MSTs. $L$ finds significance in the CST and the 20\% Threshold. The MST diameter finds no significant difference.

Fig. \ref{Scatterfig} shows that connection density is generally lower in the shape-colour binding test and the characteristic path length is generally higher. This suggests that shape-colour binding memorisation is more demanding than shape only memorisation, as would be expected since it requires inter-regional associations. 

These results point to the CST being a more sensitive method than MSTs because it keeps more information about the underlying weighted network. Further, the CST picks up the differences found at both high and low threshold levels, where different measures appear more sensitive. However, more studies must be conducted to confirm our conclusions. 

We note that if the weakest connection in an MST lies above the corresponding CST, the MST must necessarily be the same when the algorithm is applied to the CST network where information of relative strength of weights is retained as to the underlying weighted complete network. The converse is not true since two different such CST networks can have the same MST. 

\section{Conclusions}
We introduced an unbiased threshold for weighted weighted networks where many spurious low weights are prevalent. The methodology leading to this threshold also addressed important issues in understanding network topology. We tested the CST in functional brain networks for EEG data of volunteers performing cognitive tasks and found that results were equivalent to different choices of connection level thresholds and outperformed the MSTs. This threshold can be tested on other data sets to see how it compares with results reported. Particularly, we will test this in applications to the preclinical detection of Alzheimer's disease.

\addtolength{\textheight}{-12cm}   






\end{document}